\documentclass[aps,pre,twocolumn, 10pt]{revtex4}
\makeatletter
\def\@dotsep{4.5}
\makeatother
\usepackage{textcomp}
\usepackage{mathrsfs}
\usepackage{amssymb}
\usepackage{color}
\usepackage{latexsym}
\usepackage{epsf}
\usepackage{epsfig,bm}
\begin{document}

\title{Suppression of the rate of growth of dynamic heterogeneities and its relation to the local structure in a supercooled polydisperse liquid}
\author{Sneha Elizabeth Abraham}
\author{Biman Bagchi\footnote[1]{email: bbagchi@sscu.iisc.ernet.in}}
\affiliation{Solid State and Structural Chemistry Unit, Indian Institute of
Science, Bangalore 560 012, India}

\begin{abstract}
The relationship between the microscopic arrangement of molecules in a supercooled liquid and its slow dynamics at low temperature near glass transition is studied by Molecular Dynamics (MD) simulations. A Lennard-Jones liquid with polydispersity in size and mass of constituent particles is chosen as the model system. Our studies reveal that the local structure (that varies with polydispersity) plays a crucial role both in the slowing down of dynamics and in the growth of the dynamic heterogeneities, besides determining the glass forming ability (GFA) of the system. Increasing polydispersity at fixed volume fraction is found to suppress the rate of growth of dynamic correlations, as detected by the growth in the peak of the non-linear density response function, $\chi_{4}(t)$. The growth in dynamical correlation is manifested in a stronger than usual breakdown of Stokes-Einstein relation at lower polydispersity at low temperatures and also leads to a decrease in the fragility of the system with polydispersity. We show that the suppression of the rate of growth of the dynamic heterogeneity can be attributed to the loss of structural correlations (as measured by the structure factor and the local bond orientational order) with polydispersity. While a critical polydispersity is required to avoid crystallization, we find that further increase in polydispersity lowers the glass forming ability.\\

\end{abstract}
\maketitle

\section{Introduction}
      The relation between the local structure and its slow dynamics
in a supercooled liquid near glass transition temperature, $T_{g}$ is currently a subject of intense curiosity. The most distinctive feature of glass formation is the rapid increase of viscosity with decrease in temperature. The temperature at which the viscosity becomes $10^{13}$ Poise is defined as the glass transition temperature. One of the main difficulties in understanding the glass transition phenomenon is that this enormous slowing down of dynamics is apparently not accompanied by a growing static correlation length (unlike the usual critical phenomena). Static structural quantities do not reveal any long range correlation. In fact the static structure of the liquid near glass transition is not much different from its equilibrium high temperature counterpart.

      In the Adam-Gibbs picture \cite{adamgibbs}, the sharp slowing
down is related to the growth of a cooperative dynamic length scale. 
In a separate theoretical study the size of heterogeneous reconfiguring
regions in a deeply viscous liquid was inferred via the Random First Order Transition Theory (RFOT)\cite{wolynes}. There is now increasing
evidence from both experiments and simulations of a dynamic
correlation length that grows upon approaching the glass
transition \cite{glotzerNat,bertheirSci,biroliPRL}.
Multipoint susceptibilities have been devised to quantify the behavior
and magnitude of growing dynamic length scales and have been
used in the experimental studies for several materials \cite{bertheirSci}.
These have directly determined the number of molecular units that move cooperatively near glass transition. The simplest density correlation function that contains information on correlated motion is the fourth-order \cite{chandanEPL,glotzerJCP}. The four-point time-dependent density correlation function, $g_{4}(r, t)$ measure
the spatial correlations between the local liquid density at two points in space, each at two different times. The dynamical four-point
susceptibility, $\chi_{4}(t)$ (the volume integral of $g_{4}(r, t)$) becomes increasingly pronounced as glass transition is approached. 

      In this study, we look for a possible relationship between the
structure and the slowdown of dynamics in supercooled polydisperse liquids near glass transition. In particular, we look at how the local structure (which we characterize using structure factor and bond orientational order parameters) would influence the growth of dynamic heterogeneity and the glass forming ability of the system. Polydisperse liquids are one of the simplest model systems that exhibit glass transition and can be conveniently studied via both experiments \cite{WeekSc,glotzerJCP} and computer simulations as the size distribution of particles prevents crystallization \cite{murarka,sneha}. It also serves as a model for
colloids and many other real world systems like polymers, pigments, paints etc as polydispersity is inherent in all these systems. Polydispersity introduces a distribution of particle diameters and masses and thus makes the system intrinsically more heterogeneous. However, the effect of polydispersity on dynamic heterogeneity has not yet been examined in detail. Here we probe this is detail using the dynamical four-point susceptibility, $\chi_{4}(t)$. Increasing polydispersity results in the loss of structural order. Thus by varying polydispersity one can understand the effect of loss of structure on the growth of dynamic heterogeneities. Our studies \cite{sneha} have shown that increasing polydispersity at fixed volume fraction decreases the fragility. And hence this study also presents us with an opportunity to probe the growth of four-point susceptibility (and thus dynamic heterogeneity) in systems with varying degree of fragility.

      The rest of the paper is organized as follows.
In section $II$ we describe the model and
simulation details and also define various quantities that are used in the analysis. In section $III$ we present our results and give detailed
discussions on the same. We give our concluding remarks in section $IV$.

\section{Theory and computational methods}
   \subsection{Four-point susceptibility}
      The two-point, two-time, fourth-order density
correlation function \cite{chandanEPL,glotzerJCP} is defined as
\begin{eqnarray*}
g_{4}(\vec{r}_{1},\vec{r}_{2},t)& \equiv & \langle \rho(\vec{r_{1}},0) \rho(\vec{r_{1}},t) \rho(\vec{r_{2}},0) \rho(\vec{r_{2}},t) \rangle \\
& & - \langle \rho(\vec{r_{1}},0) \rho(\vec{r_{1}},t) \rangle \langle \rho(\vec{r_{2}},0) \rho(\vec{r_{2}},t) \rangle
\end{eqnarray*}
The volume integral of $g_{4} (r_{1}, r_{2}, t)$ gives the four-point susceptibility $\chi_{4}(t)$,
\begin{equation}
\chi_{4}(t) = \frac{\beta V}{N^{2}} \int \int d\vec{r_{1}} d\vec{r_{2}} 
\rho(\vec{r_{1}},0) \rho(\vec{r_{2}},t) g_{4}(\vec{r}_{1},\vec{r}_{2},t)
\end{equation}
It has been shown that $\chi_{4}(t)$ can be written as \cite{glotzerJCP}
\begin{equation}
\chi_{4}(t) = \frac{\beta V}{N^{2}} [ \langle Q^{2}(t) \rangle - \langle Q(t) \rangle ^{2} ] 
\end{equation}
Here $ \beta= \frac{1}{ k_{B}T}$ and $Q(t)$ is a time-dependent order parameter and is given by
\begin{eqnarray*}
Q(t) &=& \int \int d\vec{r_{1}} d\vec{r_{2}} \rho(\vec{r_{1}},0) \rho(\vec{r_{2}},t) w( \mid \vec{r_{1}} - \vec{r_{2}} \mid ) \\
& & = \sum_{i=1}^{N} \sum_{j=1}^{N} d\vec{r} w( \mid \vec{r_{1}} - \vec{r_{2}} \mid ) \delta(\vec{r} + \vec{r}_{i}(0) - \vec{r}_{j}(t))
\end{eqnarray*}
$w(r)$ is the overlap function that is unity inside a region of size $a$ and zero otherwise, where $a$ is taken on the order of particle diameter. In our studies we choose $a = 0.40$ for all the systems with different polydispersity. $Q(t)$ measures the number of particles that in a time $t$ has either remained within a distance $a$ of their original position (when $i = j$ ) or were replaced by another particle (when $i \ne j$ ). We can separate $Q$ into self and distinct parts, $Q(t) = Q_{S}(t) + Q_{D}(t)$.
The self part corresponds to terms with $i = j$,
$Q_{s}(t) = \sum w( \mid \vec{r_{i}(0)} - \vec{r_{j}(t)} \mid ) $.
The distinct part is given by
$Q_{D}(t) = \sum \sum_{i \ne j} w( \mid \vec{r_{i}(0)} - \vec{r_{j}(t)} \mid )) $.
The susceptibility $\chi_{4}(t)$ can then be decomposed into self, distinct
and cross terms \cite{glotzerJCP},
\begin{equation}
\chi_{4}(t) = \chi_{4}^{S}(t) + \chi_{4}^{D}(t)  + \chi_{4}^{SD}(t)
\end{equation}
where,
\begin{equation}
\chi_{4}^{S}(t) \propto \langle Q_{S}^{2}(t)\rangle - \langle Q_{S}(t) \rangle ^{2}
\end{equation}
\begin{equation}
\chi_{4}^{D}(t) \propto \langle Q_{D}^{2}(t)\rangle - \langle Q_{D}(t) \rangle ^{2}
\end{equation}
and,
\begin{equation}
\chi_{4}^{SD}(t) \propto \langle Q_{S}(t) Q_{D}(t)\rangle - \langle Q_{S}(t) \rangle \langle Q_{D}(t) \rangle                  
\end{equation}
As has been found in previous studies \cite{glotzerJCP}, we find
that for our model system also the major
contribution to $\chi_{4}(t)$ comes from $\chi_{4}^{S}(t)$ and hence
in this paper we have presented results only for $\chi_{4}^{S}(t)$.

The definition of $\chi_{4}(t)$ in Eq. (1) is in terms of spontaneous
fluctuations of local
dynamics. Berthier et al have used fluctuation dissipation theorem (FDT) and
have defined four-point susceptibilities in terms of the response
of the averaged two-time dynamical correlators to an
infinitesimal perturbing field \cite{bertheirSci,berthierJCP},
\begin{equation}
\chi_{x}(t) = \frac{ \partial \langle \mid f(t) \mid_{r} \rangle} {\partial x}
\end{equation}
where $x = T or \rho$.
Here $\partial \langle \mid f(t) \mid_{r} \rangle$ is a standard two-time
correlator and $f(r, t) = O(r, t)O(r, 0)$.
For instance when the observable $O$ is the excess density $\rho(r, t) - \rho \rho_{0}$,
then $\partial \langle \mid f(t) \mid_{r} \rangle$ is the
intermediate scattering function.
Here $ \mid f(t) \mid_{r} = \frac{1} {V} \int d\vec{r} f(\vec{r},t) $ is the
spatial average over a large but finite volume, $V$. The above defintion
of $\chi_{x}(t)$ provides a very valuable
experimental tool to measure the dynamic length scales in glass systems as shown in
\cite{bertheirSci}.

\begin{figure}
\begin{center}
\epsfig{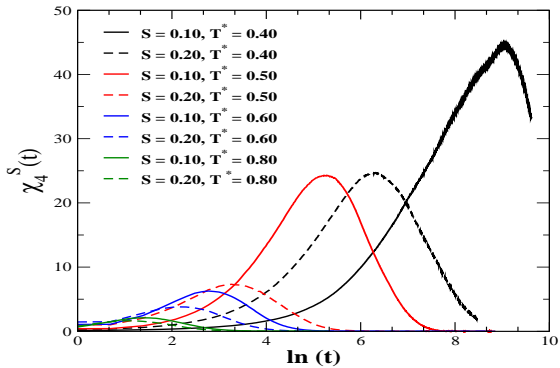}
\caption{The time dependence of the four-point susceptibility ($\chi_{4}^{S}(t)$) at four different temperatures and at two different polydispersity, $S=0.10$ (thick lines) and $S=0.20$ (dashed lines). From the bottom to the top, temperature decreases. $\chi_{4}^{S}(t)$ grows for both the systems as $T$ decreases but there is a more pronounced growth at lower polydispersity.}
 \label{chi4}
 \end{center}
 \end{figure}

\subsection{Bond-orientational order}
        The average microscopic structure of liquids is usually
described by the radial distribution function or the structure factor, which essentially measures only the density-density correlation function.
However, bond-orientational order parameters ($BOP$) introduced
by Steinhardt et al \cite{steinhardt,wang1,wang2} gives a better
quantification of the local structure as they capture the symmetry
of bond orientations. BOP is described in terms of combinations of
spherical harmonic functions. Consider a system of $N$ particles.
First, one defines a set of \textasciigrave bonds\textasciiacute which are defined as the vectors connecting neighboring particles. All particles {\em j} within a cutoff distance $r_{0}$ of particle {\em i} are defined as neighbors of particle {\em i}. Here $r_{0}$ is chosen to be equal to the distance to the first minimum of the radial distribution function ($RDF$). The local order parameters associated with a bond $r$ are the set of numbers
\begin{equation}
Q_{lm}(r ) \equiv Y_{lm}(\theta(r), \phi(r) ) 
\end{equation}
where $\theta(r)$ and $\phi (r)$ are the polar and azimuthal angles of the bond with respect to an arbitrary but fixed reference frame and
$Y_{lm}(\theta(r), \phi(r))$  are the spherical harmonic functions.
It is useful to consider only the even-l spherical harmonics, which
are invariant under inversion. Global bond order parameters
can be calculated by averaging over all the bonds in the system,
\begin{equation}
\overline Q_{lm} \equiv \frac{1}{N_{b}} \sum_{bonds} Q_{lm}(\vec{r})
\end{equation}
Since $Q_{lm}’s$ for a given $l$ depends on the rotations of the reference frame, it is important to consider the rotationally invariant combinations such as
\begin{equation}
 Q_{l} \equiv [\frac{4 \pi}{2l+1} \sum_{-l}^{l} \mid \overline  Q_{lm} \mid^{2}]^{\frac{1}{2}}
\end{equation}
and
\begin{equation}
 W_{l} \equiv  \sum_{m_{1},m_{2},m_{3}}^{m_{1}+m_{2}+m_{3}=0}(.....) \overline Q_{lm_{1}} \overline Q_{lm_{2}}\overline Q_{lm_{3}} 
\end{equation}
$Q_{l}$ and $W_{l}$ are the second and third order invariants, respectively. The coefficients $(...)$ are Wigner 3j symbols. One also defines a normalized quantity,
\begin{equation}
\hat{W_{l}} \equiv  \frac{W_{l}}{(\sum_{m} \mid Q_{lm} \mid^{2})^{3/2}}
\end{equation}
which for a given $l$ is independent of the magnitudes of the $Q_{lm}$.
The four bond order parameters $Q_{4}$, $Q_{6}$, $\hat{W_{4}}$ and Ŵ$\hat{W_{6}}$ are sufficient to identify different crystal structures. The typical values of these for different crystal structures are given in \cite{wang2}. For a liquid the global values of all these four quantities are zero as there is no long range order. Note that in clusters with cubic symmetry non-zero averages occur only for $l \geq 4$ whereas non-zero averages occur only at $l=6$ and $10$ for icosahedral cluster.

\begin{figure}
\begin{center}
\epsfig{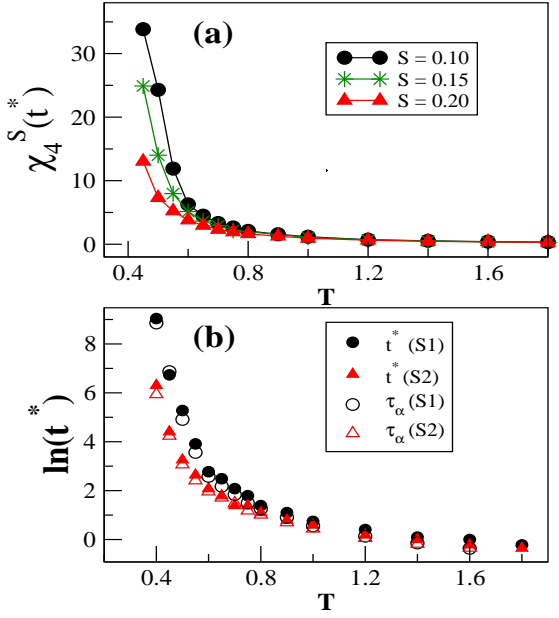}
\caption{ {\bf(a)} The value of the peak height of $\chi_{4}^{S}(t)$ is plotted as a function of $T$ for $S=0.10$ (filled circle), $S=0.15$ (star) and $S=0.20$ (filled triangle) systems. The figure shows the suppression of the rate of growth of dynamic heterogeneity with $S$.
{\bf(b)} The time at which $\chi_{4}^{S}(t)$ peaks, $t^{*}$ is plotted as a function of $T$ for $S=0.10$ (S1) and $S=0.20$ (S2) systems (filled circles and triangles, respectively). $t^{*}$ is similar to the $\alpha$-
relaxation time (open circles and triangles, respectively).
The latter is obtained by doing KWW
fit to $F_{s}(k_{max}, t)$ where $k_{max}$ corresponds to the first peak in the static structure factor.}
 \label{chi4max}
 \end{center}
 \end{figure}

 \subsection{System and simulation details}
      Micro canonical (NVE) ensemble MD simulations are carried out in
three dimensions on a system of $N = 864$ particles of mean diameter $\overline \sigma$ with polydispersity in both size and mass. The polydispersity in size is introduced by random sampling from the Gaussian distribution of particle diameters $\sigma$,
\begin{equation}
P(\sigma) = \frac{1}{\sqrt{2 \pi}\delta}exp[-\frac{1}{2}(\frac{\sigma - \overline \sigma} {\delta})^{2}]
\end{equation}
The standard deviation $\delta$ of the distribution divided by its mean $ \overline \sigma$ gives a dimensionless
parameter, the polydispersity index $S = \frac {\delta} {\overline \sigma}$. The mass $m_{i}$ of particle $i$ is scaled by its diameter $m_{i} = \overline m(\frac{\sigma_{i}}{\overline \sigma})^3$. ⎞
We have chosen $\overline m=1.0$. 
The simulations are carried out at different values of the polydispersity index, S but at fixed volume fraction, $\phi = 0.52$.
The interactions between the particles are given by the
shifted-force Lennard-Jones (LJ) 12-6 potential
\begin{equation}
U_{ij} = 4\epsilon_{ij}[ ( \frac{\sigma_{ij}}{r_{ij}} )^{12} -  ( \frac{\sigma_{ij}}{r_{ij}} )^{6} ]
\end{equation}
where $i$ and $j$ represent any two particles and
$\sigma_{ij} = ( \frac{\sigma_{i}+\sigma_{j}}{2})$. 
The LJ interaction parameter $\epsilon_{ij}$ is assumed to be 
the same for all particle pairs and set equal to unity.
The particles are enclosed in a cubic box and periodic boundary conditions are used. The cutoff radius $r_{c}$ is chosen to be $2.5 \overline \sigma$.
A time step of $0.001$ is employed for $T \geq 1.0$ and $0.002$ for $T < 1.0$. All quantities in this study are given in reduced units (length in units of $\sigma$, temperature in units of $\frac{\epsilon}{k_{B}}$ and time in units of $\tau=( \frac{\overline m  \overline \sigma^{2}}{\epsilon } ) ^{\frac{1}{2}}$).

\begin{figure}
 \begin{center}
 \epsfig{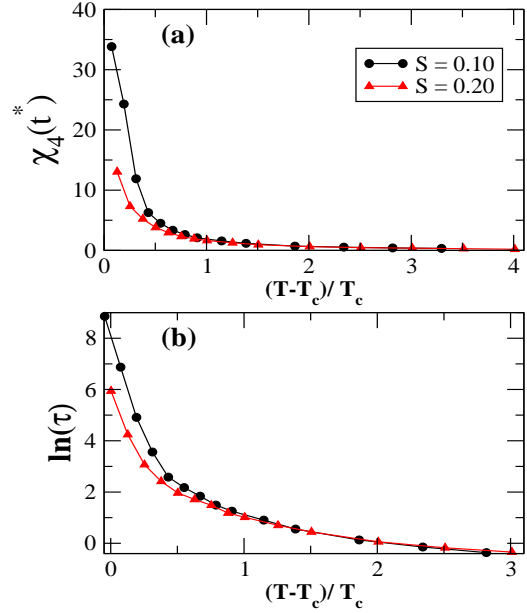}
 \caption{{\bf (a)} $\chi_{4}^{S}(t^{*})$
as a function of the distance from the MCT critical temperature $T_{c}$
for $S=0.10$ and $S=0.20$ systems. $T_{c}$ values are $0.42$ and $0.39$ for $S=0.10$ and $S=0.20$ systems, respectively.{\bf (b)} $\alpha$-
relaxation time, $\tau_{\alpha}$ as a function of the distance from $T_{c}$.}
 \label{chi4rescale}
 \end{center}
 \end{figure}

 \section{Results and discussion}
      The main objective of our study is to demonstrate the effect of
polydispersity and hence of the local structure on the growth of dynamic heterogeneities. By varying polydispersity we can \textasciigrave tune\textasciiacute the local structure and hence study its effects on the dynamic heterogeneity. As polydispersity is increased, the local structure is progressively destroyed. Hence the blocking of the particles in the cages of the neighboring particles (as required for the mode coupling theory of dynamic transition \cite{leutheusser}) becomes ineffective at higher polydispersity. We find that this has a pronounced effect on the development of dynamic heterogeneities as well. In this section we systemically present our results and show that the local structure plays a very important role in determining the dynamics in supercooled liquids near glass transition.

\begin{figure}
 \begin{center}
 \epsfig{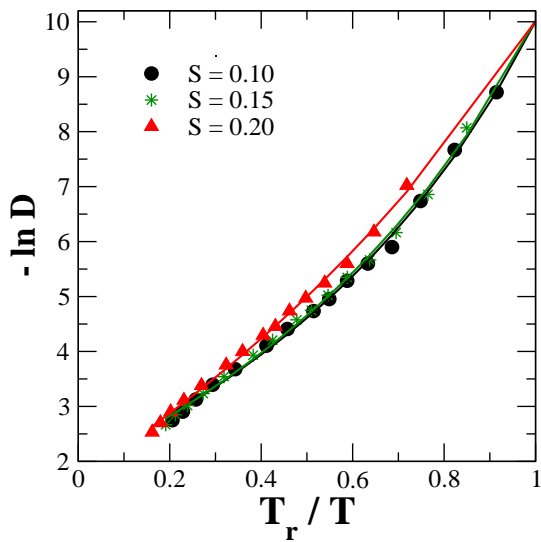}
 \caption{Angell-like fragility plot at different $S$. The thick lines are VFT fit to the diffusivity data, $D=D_{0}exp(\frac{E_{D}}{T-T_{0}})$.
The reference temperature $T_{r}$ is chosen such that $D(T_{r})=4.5 \times 10^{-5}$. The VFT extrapolation is used to locate $T_{r}$. The plot shows that fragility decreases with $S$. Strength parameter $m$
(where $m=\frac{E_{D}}{T_{0}}$ \cite{Angell}) obtained from VFT fit have the values $7.78$, $8.54$ and $15.94$ for $S=0.10$, $S=0.15$ an
d $S=0.20$ systems, respectively.}
 \label{fragility}
 \end{center}
 \end{figure}

       \subsection{Suppression of the rate of growth of dynamic
                correlations by polydispersity}

      The four-point susceptibility $\chi_{4}^{S}(t)$  obtained from
Eq. (2) is shown in FIG. \ref{chi4} for $S=0.10$
and $S=0.20$ for a few temperatures. From Eqs. (1) and (2) we see that $\chi_{4}(t)$ becomes larger when the dynamic fluctuations become increasingly spatially correlated. Since $\chi_{4}(t)$ is the volume integral of the four-point correlator $g_{4}(r, t)$, it is directly related to the number of correlated particles. As temperature is lowered $\chi_{4}^{S}(t)$ grows for both the systems but the rate of growth decreases with polydispersity. This more clearly seen in FIG. \ref{chi4max}(a) where the peak height of $\chi_{4}^{S}(t)$ (which we label as $\chi_{4}^{S}(t^{*})$) is plotted against temperature for different values of polydispersity.
FIG. \ref{chi4max}(a) shows the suppression of the rate
of growth of dynamical heterogeneity by polydispersity. In 
FIG. \ref{chi4rescale}(a) we plot the peak height $\chi_{4}^{S}(t^{*})$ 
as a function of the distance from the MCT critical temperature $T_{c}$
for $S=0.10$ and $S=0.20$ systems. The rescaled plot also shows a
suppression in the rate of growth of $\chi_{4}^{S}(t^{*})$ for $S=0.20$ system as compared to $S=0.10$ system. 
By fitting to the expression
$\chi_{4}(t^{*}) \sim (\frac{T-T_{c}}{T_{c}})^{\gamma_{\chi}}$, we get the
values of the exponent $\gamma_{\chi}$ as $1.507$ and $1.188$ for
$S=0.10$ and $S=0.20$ systems, respectively.
The exponent $\gamma_{\chi}$ thus seems to change with
polydispersity. The suppression of the rate of growth of dynamical heterogeneity by polydispersity leads to the dynamic cross-overs observed in the values of the stretch exponent, $\beta$ and the non-Gaussian parameter, $\alpha_{2}(t)$ between $S=0.10$ and $S= 0.20$ systems as shown earlier \cite{sneha} and the cross-over behavior seen in the exponent $z_{\sigma}$ that quantifies the deviation from the prediction of Stokes-Einstein relation(See Section III. B).
Increasing polydispersity at
fixed volume fraction decreases the fragility of the system
(See FIG. \ref{fragility}). Fragility measures the rapidity with which the system approaches glass transition. Hence decrease in the 
the rate of growth of dynamic heterogeneity with polydispersity 
is consistent with the decrease in the fragility. This is further explained in section III.C.

	In FIG. \ref{chi4max}(b) we show the time at which
$\chi_{4}^{S}(t)$ peaks, $t^{*}$ versus temperature for $S=0.10$ and $S=0.20$. Also shown are the $\alpha$-relaxation times, $\tau_{\alpha}$ obtained by doing Kohlrausch-William-Watts (KWW) fit to the self part of
intermediate scattering function, $F_{s}(k, t)$. The plot shows that
the dynamics is maximally correlated on time scales of the order of the alpha relaxation time. In FIG. \ref{chi4rescale}(b), $\tau_{\alpha}$ is plotted as a function of the distance from the MCT critical temperature, $T_{c}$. The rescaled plot shows that the rate of growth of $\tau_{\alpha}$
decreases with polydispersity.

 \begin{figure}
 \begin{center}
 \epsfig{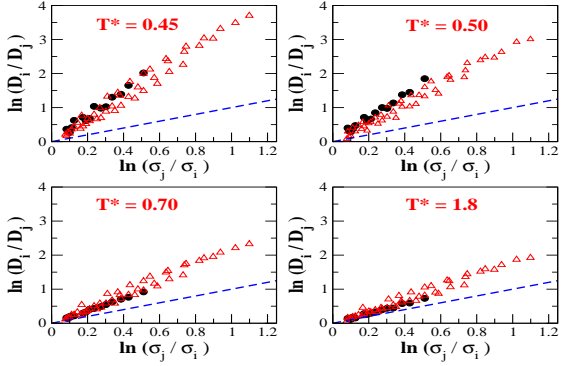}
 \caption{The plot of $ln(D_{i}/D_{j})$ versus $ln(\sigma_{j}/\sigma_{i})$ for $S=0.10$(circles) and $S=0.20$ (triangles). Here the subscript $i$ denotes the smaller particle. If the $SE$ relation is valid, this plot would be of unit slope (dashed line). The plot shows that at high temperature the deviation from the Stokes-Einstein prediction is higher for $S=0.20$ system, but the
scenario reverses at low temperature where $S=0.10$ system shows a stronger deviation due to the faster growth of dynamic heterogeneity.}
 \label{SElog}
 \end{center}
 \end{figure}

 \subsection{Breakdown of Stokes-Einstein Relation}
\begin{figure}
 \begin{center}
 \epsfig{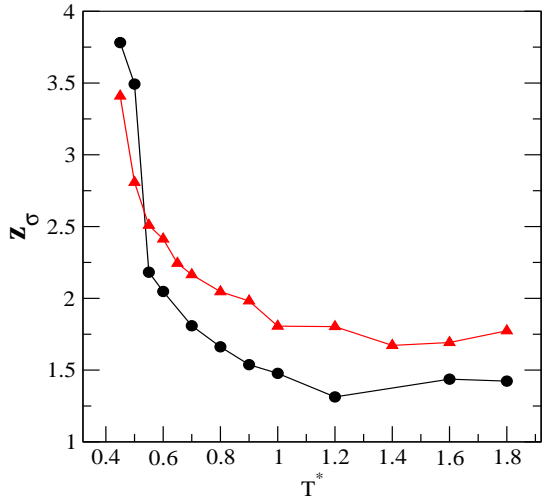}
 \caption{ The values of the power law exponent, $z_{\sigma}$ obtained by fitting Eq. 16, are plotted against $T$, for $S=0.10$ and $S=0.20$ systems. If the $SE$ relation is strictly valid then $z_{\sigma}=1$. Deviation from unity shows the breakdown of $SE$ relation. The figure clearly shows that at high temperatures
intrinsic heterogeneity causes a larger breakdown in $S=0.20$ system, whereas at low temperatures the faster growth of dynamic heterogeneity leads to a stronger breakdown in $S=0.10$ system.
 }
 \label{SEexpo}
 \end{center}
 \end{figure}

      In this section we discuss the breakdown
of Stokes-Einstein ($SE$) relation and its connection to the rate of growth of dynamic heterogeneity in the system as temperature is lowered. The $SE$ relation is based on treating the liquid as a continuum and is given by,
\begin{equation}
  D=\frac{k_{B}T}{C \eta \sigma} 
\end{equation}
Here $C$ is a constant that depends on the boundary conditions (stick or slip) and $\eta$ is the viscosity. If the Stokes-Einstein relation
is strictly valid, then a plot of $ln(D_{i}/D_{j})$ versus $ln(\sigma_{j}/\sigma_{i})$ would be a straight line with unit slope. Here $i$ and $j$ are indices for solute and solvent, respectively. In FIG. \ref{SElog} we show this plot for $S=0.10$
and $S=0.20$ systems. Both systems show deviation from the SE
prediction even at high temperatures due to the intrinsic heterogeneity
in the system, with the deviation being more pronounced for $S=0.20$ system. The SE relation has been shown to be not
valid for the diffusion of small solutes in a solvent of bigger particles \cite{sarikaJCP}. There is an anomalous enhancement of the self-diffusion over the SE value for small solutes which can be described by a power law,
\begin{equation} 
  \frac{D_{i}}{D_{j}} \sim (\frac{\sigma_{j}}{\sigma_{i}})^{z_{\sigma}}
\end{equation}
Hence the exponent $z_{\sigma}$ quantifies the deviation from $SE$ relation. It is unity in the $SE$ limit and usually larger than unity in supercooled liquid. FIG. \ref{SEexpo} shows that $z_{\sigma}$ deviates significantly from unity for polydisperse liquids, particularly at low temperatures. Interestingly, it is larger than unity even at high temperature because of the heterogeneity in the size and mass. This deviation of the slope from unity at high temperature can be a combination of two different effects. The first one is the mass which is not present in SE relation but has been reported earlier in simulations \cite{barrat} and mode coupling theory (MCT) studies \cite{sarikaPRE}. The studies predict a weak power law mass dependence of diffusion. The second effect
is that of size which has also been obtained in experiments and simulations \cite{pollack} and MCT studies \cite{sarikaJCP}. When the size of one of the particles is $1.5-15$ times smaller than the other it shows an anomalous rise in diffusion. This enhanced diffusion has been explained in terms of microviscosity effect. The MCT studies explain the microscopic origin of the size effect in terms of the difference in relaxation timescales of the two particles which leads to a decoupling in the dynamics \cite{sarikaJCP}. (Note that for size dependent studies the small particle was a tracer and for mass dependent studies the heavier particle was a tracer. In the present study the systems are intrinsically heterogeneous.)

      As temperature is lowered the deviation from the prediction of
$SE$ relation becomes more pronounced and one observes a crossover in the value of $z_{\sigma}$ between the $S=0.10$ and $S=0.20$ systems; the values are much higher for $S=0.10$ system than $S=0.20$ at low temperatures [see FIG. \ref{SEexpo}]. This again shows that the rate of growth of dynamic heterogeneity is faster in $S=0.10$ system than in $S=0.20$ system. The faster rate of growth of dynamic correlations leads to similar temperature-dependent crossovers between $S=0.10$ and $S=0.20$ systems in the values of the stretch exponent ($\beta$) and the non-Gaussian parameter ($\alpha_{2}(t)$), both of which contain implicit information on dynamic heterogeneity \cite{sneha}. These studies show that there is a strong correlation between the growing dynamic heterogeneity in the system and the breakdown of $SE$ relation as the former renders the continuum description of liquid invalid as required for the $SE$ relation.

 \subsection{Fragility and the growth of dynamic correlations}
      It has been shown that the dynamics of fragile liquids are more
spatially heterogeneous than that of strong liquids \cite{bohmer}. Increasing polydispersity at fixed volume fraction decreases the fragility of the system as shown in FIG. \ref{fragility}. The decrease of the rate of growth of $\chi_{4}(t)$ with polydispersity supports the previously observed correlation between fragility and dynamic heterogeneity. The intrinsic heterogeneity of
the system (as measured by the distribution of particle masses and sizes)
increases with polydispersity. Hence we have the interesting scenario in which increasing polydispersity leading to a more {\em homogeneous} dynamics even though the system becomes completely amorphous at higher values of polydispersity. It has been shown that polydispersity has a pronounced effect on potential energy surface \cite{lacks}. As polydispersity is increased from zero, the characteristics of the potential energy minima change from that of crystalline to that of amorphous. The latter is known to have low curvature and small barriers along some coordinates \cite{bairdPRL,huerPRL}. This observation is also consistent from the perspective of the inherent structure formalism, according to which the potential energy landscape of a fragile liquid is very heterogeneous which in turn leads to heterogeneous dynamics whereas the landscape of strong liquids consist of a single mega basin \cite{stillinger}. Hence from a potential energy landscape perspective, increasing polydispersity leads to a {\em smoothening} of the landscape that in turn leads to the facilitation of dynamics as well as decrease of fragility. In Section III. D, we try to understand how polydispersity suppresses the rate of growth of dynamic correlations and in particular, whether the loss of structure upon increasing polydispersity has any role to play in this. 

\begin{figure}
 \begin{center}
 \epsfig{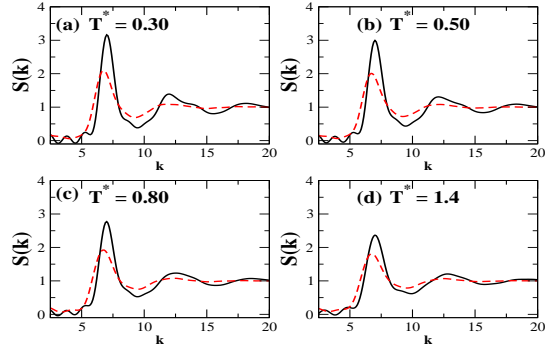}
 \caption{The calculated static structure factor, $S(k)$ is plotted against wave number $k$ for $S=0.10$
(thick lines) and $S=0.20$ (dashed lines) systems at a few temperatures.
Structural correlation is weaker in $S=0.20$ system than in $S=0.10$ system and shows no appreciable change with temperature. }
 \label{SK}
 \end{center}
 \end{figure}

\begin{figure}
 \begin{center}
\epsfig{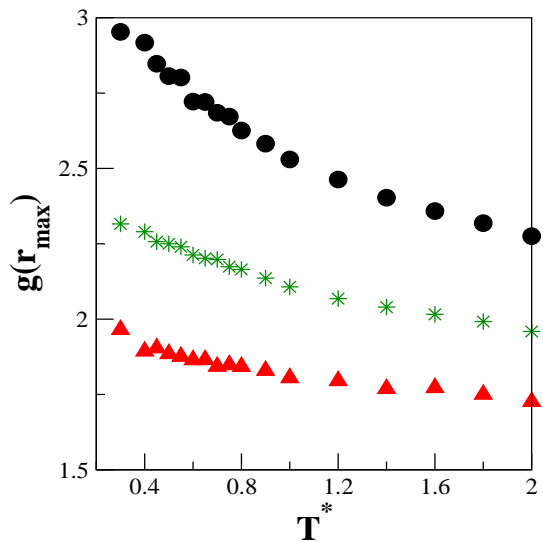}
 \caption{Peak height value of the radial distribution function, $g(r_{max})$, when plotted against separation $r$, is plotted as a function of $T$ for $S = 0.10$ (circles), $S = 0.15$ (stars) and $S = 0.20$ (triangles). $S=0.20$ system does not show any remarkable change in the value of
$g(r_{max})$ upon lowering of $T$. On the other hand, $S=0.10$ system shows a sudden increase of spatial correlations
for $T \leq 0.8$. Comparison between FIG. \ref{grmax} and FIG. \ref{chi4max}(a) shows that local structure plays a
crucial role in the build up of dynamic correlations.}
 \label{grmax}
 \end{center}
 \end{figure}

\subsection{Local structure and the growth of dynamic heterogeneities}

      We plot the static structure factor, $S(k)$ for $S=0.10$ and
$S=0.20$ systems in FIG. \ref{SK}. The plot shows that increasing polydispersity destroys the local structure in the system
as the system becomes more amorphous. The peak height of $S(k)$ is highly suppressed in $S=0.20$ system as compared to $S=0.10$ system and does not show any appreciable growth upon lowering of $T$. FIG. \ref{grmax} shows the peak height value of $RDF$, $g(r_{max})$ as a function of temperature. At $S=0.10$, the peak height shows considerable enhancement upon lowering of temperature whereas at $S=0.20$ there is no remarkable change in the value of $g(r_{max})$ with temperature.

      As mentioned in Section II. B, the bond orientational order
parameters give a better quantification of the local structural arrangement. Frank \cite{frank} proposed that atoms might form icosahedral clusters in liquids since the lowest energy state of a $13$-atom cluster interacting via Lennard-Jones potential is an icosahedron (and not fcc). But icosahedra cannot tile space in $3$-dimensions due to its $5$-fold symmetry and hence do not satisfy the global structural stability criterion. This geometrical frustration could be an important factor that contributes to the stability of glassy state \cite{egami}. Steinhardt et al \cite{steinhardt}
have shown that there is a long range orientational icosahedral
order in supercooled liquids. It has been shown that the large size disparities at higher values of polydispersity would inhibit any icosahedral cluster formation. However, since at low/moderate polydispersity the peak height of $g(r_{max})$
shows a pronounced growth as temperature is lowered, one can ask
whether this is due to the formation of icosahedral clusters that
grows with decrease in temperature.

\begin{figure}
\begin{center}
\epsfig{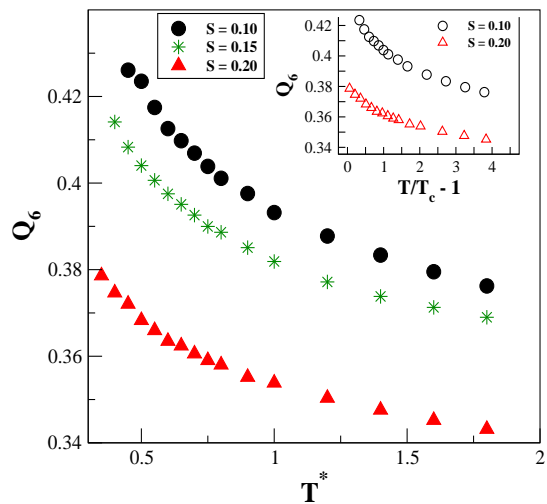}
\caption{The calculated local values of bond order parameter, $Q_{6}$ are plotted as a function of $T$ for $S = 0.10$ (circles), $S = 0.15$ (stars) and $S = 0.20$ (triangles). The plot shows that at the local level there is significant orientational order that increases with decrease in temperature and decreases with polydispersity.[{\bf Inset}: $Q_{6}$ as a function of the distance from $T_{c}$.]}
 \label{Q6loc}
 \end{center}
 \end{figure}

      We look for the local values of $BOP$ in order to understand
whether local orientational order plays any role in the growth of dynamic heterogeneities. The icosahedral order, if present, would be picked by the $BOP$ corresponding to $l=6$, $Q_{6}$. The local values of $Q_{6}$ are plotted in FIG. \ref{Q6loc}. To get the local values the spherical harmonics corresponding to $l=6$ are summed over the nearest neighbor bonds only. The figure shows that there is a pronounced icosahedral orientational order at the local level. This local icosahedral order shows considerable enhancement at lower polydispersity as temperature is lowered (for $T \leq 0.80$) and also decreases with polydispersity. 
In the inset of FIG. \ref{Q6loc} we plot $Q_{6}$ as a function of the distance from the MCT critical temperature $T_{c}$ which shows that as temperature
is lowered, $Q_{6}$ increases much more sharply at lower values of $S$.

\begin{figure}
 \begin{center}
 \epsfig{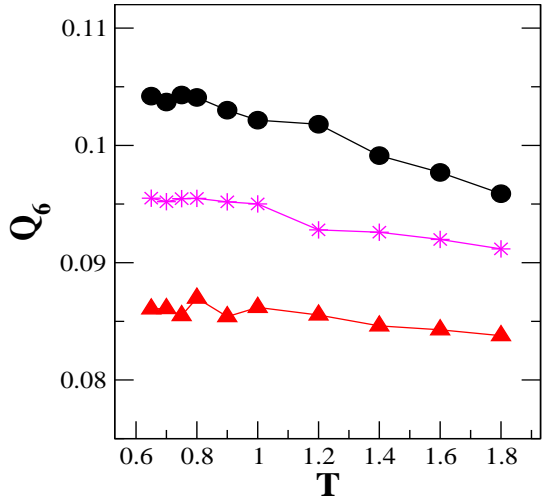}
 \caption{The calculated values of the global bond order parameter, $Q_{6}$ are plotted as a function of
$T$ for $S = 0.10$ (circles), $S = 0.15$ (stars) and $S = 0.20$ (triangles). The plot shows that there is no appreciable long range orientational order developing in the supercooled state.}
 \label{Q6glob}
 \end{center}
 \end{figure}

      In FIG. \ref{Q6glob} we plot the global values of $Q_{6}$ for
different $S$ as a function of temperature. The averages over bonds are evaluated by summing over all bonds lying within a sphere of radius $2.4$ units. Nine such spheres are considered whose centers lie at different
locations of the simulation box. We repeat this averaging for several
different snapshots obtained from simulation. It is evident from FIG. \ref{Q6glob} that polydispersity suppresses long range orientational order.
Even at moderate polydispersity, there is no pronounced growth
of long range icosahedral order upon supercooling.

\begin{figure}
 \begin{center}
 \epsfig{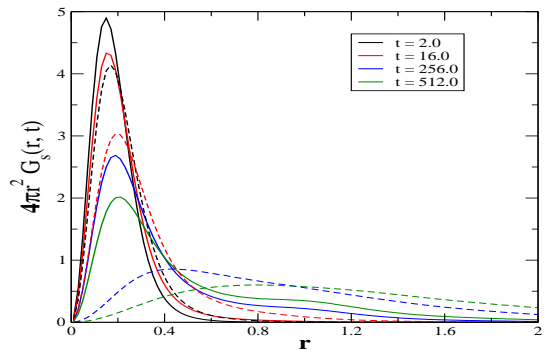}
 \caption{The calculated van Hove self-correlation function, $G_{s}(r, t)$, plotted against position $r$ at different times (indicated on the figure) for $S=0.10$ (thick lines) and $S=0.20$ (dashed lines) systems at $T=0.45$. The figure shows the faster decay of density
correlations for $S=0.20$ system as compared to $S=0.10$ system.}
 \label{vanhovSelf}
 \end{center}
 \end{figure}

\begin{figure}
 \begin{center}
 \epsfig{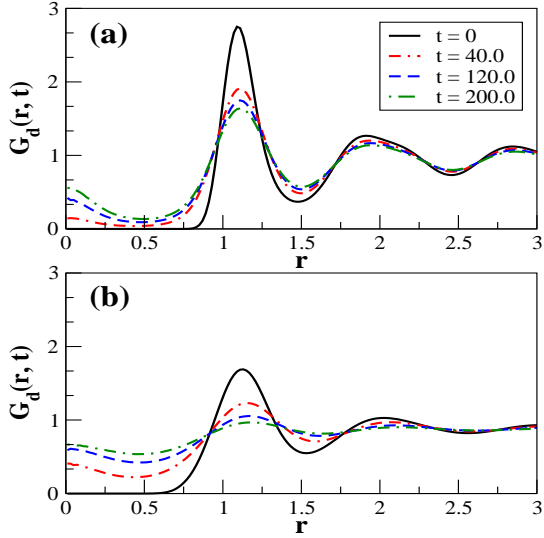}
 \caption{The calculated van Hove distinct-correlation function, $G_{d}(r,t)$ is plotted against position
$r$ for $S=0.10$ (upper panel) and $S=0.20$ (lower panel) systems, at $T=0.45$ depicting the faster decay of inter-particle correlations at higher polydispersity.}
 \label{vanhovDis}
 \end{center}
 \end{figure}

      Our results indicate that increasing polydispersity destroys both
the local structure and the local orientational order.
The four-point susceptibility $\chi_{4}^{S}(t)$ measures the susceptibility
arising from the number of localized particles and is a measure
of the dynamic heterogeneity in the system. Thus the dynamic heterogeneity
is associated with the temporary localization of particles by their neighbors. Since the local structure is destroyed, at higher values of polydispersity it is not possible to have such a caging effect. As a consequence
particle motion gets decorrelated over much shorter time scales.
This is best seen by plotting the van Hove correlation function [see
FIG.s \ref{vanhovSelf} and \ref{vanhovDis}].
Thus the loss of local structure due to polydispersity suppresses the growth of dynamic heterogeneity in the system.

  \subsection{ Polydispersity and Glass Forming Ability (GFA)}

   We find that the present system of $LJ$ particles of varying size and
mass crystallizes when polydispersity is less than $7.5 \%$. Thus, our system with $10 \%$ polydispersity can be regarded as the system on the lower side of polydispersity that could be made to avoid crystallization and remain liquid within our MD simulation time range. Interestingly, we find that this is also the system that shows glassy behavior at the highest temperature. When we increase polydispersity beyond $10\%$, we need to lower the temperature to capture the onset of slow glassy dynamics. This can be seen from FIG. \ref{chi4max}(b) which shows that the rate of growth
of $\tau_{\alpha}$ decreases with $S$.
The new aspect revealed in the present work
is the correlation between the $GFA$ and the rate of growth of the dynamic
heterogeneity \textemdash sharper the growth, larger the $GFA$.
 
      Given that a polydisperse liquid with low polydispersity ($S < 0.05$)
crystallizes easily, the loss of local structure at large $S$ ($\geq 0.20$) and the concomitant difficulty of glass formation at large $S$ imply
a rather narrow range of $S$ for polydisperse systems to act as good glass formers. This means that only at moderate polydispersity the system has a high $GFA$. The $GFA$ decreases with polydispersity beyond a value of $S$. Further insight can be gained from the study of inherent structures. We find that the ruggedness of potential energy landscape decreases with $S$, which is consistent with decrease of the $GFA$ as well as fragility with $S$. It is important to note that network glass formers like Silica which is a strong liquid in Angell’s fragile/strong classification, exhibits high glass forming ability due to trapping by defects. This apparently contradicts the decrease of $GFA$ with polydispersity. The latter appears to be a hallmark of polydisperse systems. We shall address these issues in detail elsewhere.

\section{ Concluding Remarks}
The hypothesis that {\em structure determines dynamics} has been termed by Harowell as the {\em central dogma of glass science}\cite{harowell}.
This dogma is validated in the Mode Coupling Theory.
The Adams-Gibbs theory, however, gives
larger emphasis on the emergence of a dynamical correlation length as
the source of slow dynamics which does not seem to depend too sensitively
on the structure formation. This can be understood from the relative
insensitivity of the structure to temperature. In the present work, we have
varied polydispersity that allows large variation of the local structure
and find that the local structure indeed plays an important role in the
development of dynamic correlations and the slow dynamics near glass transition in a supercooled polydisperse liquid. Increasing polydispersity at constant volume fraction leads to a suppression of the rate of growth of dynamic heterogeneity in the system, which can be attributed to the loss of local structure with polydispersity. At moderate polydispersity, there is a faster growth of structural correlations as the temperature is lowered, which leads to a corresponding faster growth of dynamic heterogeneity. At higher polydispersity, structural correlations
are weak and do not show any significant change with temperature
and correspondingly, the rate of growth of dynamic correlations is also less. We also find that there is a pronounced local icosahedral order which increases with cooling and decreases with polydispersity. No significant long range icosahedral order is found either in the equilibrium or supercooled liquid.

    An important outcome of the present work is the hitherto unknown 
correlation between polydispersity and glass forming ability. This correlation deserves further study.


\begin{thebibliography}{99}

\bibitem{adamgibbs} G. Adam and J. H. Gibbs, J. Chem. Phys. {\bf 43}, 139 (1965).

\bibitem{wolynes} X. Xia and P. G. Wolynes, Proc. Nat. Acad. Sci. {\bf 97}, 2990 (2000); Phys. Rev. Lett. 86, 5526 (2001).

\bibitem{glotzerNat} C. Bennemann, C. Donati, J. Baschnagel and S. C. Glotzer, Nature {\bf 399}, 246 (1999)

\bibitem{bertheirSci} L. Berthier et al, Science {\bf 310}, 1797 (2005).

\bibitem{biroliPRL} O. Dauchot and G. Marty, G. Biroli, Phys. Rev. Lett. {\bf 95}, 265701 (2005).

\bibitem{chandanEPL} C. Dasgupta, A. V. Indrani, S. Ramaswamy and M. K. Phani, Europhys. Lett. 15, 307 (1991).

\bibitem{glotzerJCP} S. C. Glotzer, V. N. Novikovand T. B. Schroder, J. Chem. Phys. {\bf 112}, 509 (2000); N. Lacevic et al, J. Chem. Phys. 119, 7372 (2003).

\bibitem{WeekSc} E. R. Weeks et al, Science {\bf 287}, 627 (2000).

\bibitem{CipelletiNatPhys} P. Ballesta, A. Duri and L. Cipelletti, Nature Phys. {\bf 4}, 550 (2008).

\bibitem{murarka} R. K. Murarka and B. Bagchi, Phys. Rev. E 67, 051504 (2003).

\bibitem{sneha}   S. E. Abraham, S. M. Bhattacharyya and B. Bagchi, Phys. Rev. Lett. {\bf 100}, 167801 (2008).

\bibitem{berthierJCP} L. Berthier et al, J. Chem. Phys. {\bf 126}, 184503 (2007); J. Chem. Phys. {\bf 126}, 184504 (2007)

\bibitem{steinhardt} P. J. Steinhardt, D. R. Nelson and M. Ronchetti, Phys. Rev. B {\bf 28}, 784 (1983).

\bibitem{wang1} Y. Wang and C. Dellago, J. Phys. Chem. B {\bf 107}, 9214 (2003)

\bibitem{wang2} Y. Wang, S. Teitel and C. Dellago, J. Chem. Phys.{\bf 122}, 214722 (2005).

\bibitem{Angell} C. A. Angell, J. Phys. Chem. Solids {\bf 49}, 863 (1988); C. A. Angell, Science {\bf 267}, 1926 (1995).

\bibitem{leutheusser} E. Leutheusser, Phys. Rev. A {\bf 29}, 2765 (1984); U. Bengtzelius, W. Gotze and A.Sjolander, J. Phys. C {\bf 17}, 5915 (1984).

\bibitem{sarikaJCP} S. Bhattacharyya and B. Bagchi, J. Chem. Phys. {\bf 106}, 1757 (1997).

\bibitem{barrat} F. O. Kaddour and J. L. Barrat, Phys. Rev. A {\bf 45}, 2308 (1992).

\bibitem{sarikaPRE} S. Bhattacharrya and B. Bagchi, Phys. Rev. E {\bf 61}, 3850 (2000).

\bibitem{pollack} G. L. Pollack and J. J. Enyeart, Phys. Rev. A {\bf 31}, 980 (1985); G. L. Pollack et al, J. Chem.Phys {\bf 74}, 2 (1981); G. Heuberger and H. Sillescu, J. Phys. Chem. {\bf 100}, 15255 (1985).

\bibitem{bohmer} R. Bohmer et al, J. Chem. Phys {\bf 99}, 4201 (1993); C. M. Roland and K. L. Ngai, Macromolecules {\bf 24}, 5315 (1991); K. Niss et al, J. Phys. Condens. Matter {\bf 19}, 076102 (2007).

\bibitem{lacks} D. L. Lacks and J. R. Wienhoff, J. Chem. Phys. {\bf 111}, 396 (1999).

\bibitem{bairdPRL} B. B. Baird and H. R. Schober, Phys. Rev. Lett. {\bf 66}, 9636 (1991).

\bibitem{huerPRL} A. Huer and R. J. Silbey, Phys. Rev. Lett. {\bf 70}, 3911 (1993).

\bibitem{stillinger} F. H. Stillinger, Science {\bf 267}, 1935(1995); P. B. Debenedetti and F. H. Stillinger, Nature {\bf 410}, 66, 639 (2001).

\bibitem{frank} F. C. Frank, Proc. R. Soc. London Ser. A {\bf 215}, 43 (1952).

\bibitem{egami} T. Egami in \textgravedbl Rapidly Solidified Alloys\textacutedbl, Edited by Howard H. Liberman (CRC Press, 1993).

\bibitem{harowell} P. Harowell, \textgravedbl Conference on Unifying Concepts in Glassy Physics III, Bangalore, 2004.\textacutedbl

\end{thebibliography}
\end{document}